\documentclass{article}
\usepackage{cite}
\usepackage[utf8]{inputenc}
\usepackage{amsmath}
\usepackage{amssymb,amsfonts}
\usepackage{graphicx}
\usepackage{textcomp}
\usepackage{xcolor}
\usepackage{caption} 
\usepackage{subcaption}
\usepackage{float}
\DeclareUnicodeCharacter{0394}{$\Delta$}
\DeclareUnicodeCharacter{2009}{\,} 
\def\BibTeX{{\rm B\kern-.05em{\sc i\kern-.025em b}\kern-.08em
		T\kern-.1667em\lower.7ex\hbox{E}\kern-.125emX}}
\graphicspath{ {images/} }

\begin{document}
	
	\title{Inverse Design of Potential Singlet Fission Molecules using a Transfer Learning Based Approach}
	
	\author{Akshay Subramanian\textsuperscript{1}, Utkarsh
		Saha\textsuperscript{2}, Tejasvini Sharma\textsuperscript{2},\\ Naveen K. Tailor\textsuperscript{2}, Soumitra Satapathi\textsuperscript{2{*}}}
	
	\maketitle
	
	\noindent1. Department of Metallurgical and Materials Engineering, Indian Institute of Technology Roorkee, Roorkee, India \\
	2. Department of Physics, Indian Institute of Technology Roorkee, Roorkee, India \\
	{*}corresponding author:
	Soumitra Satapathi (soumitrasatapathi@gmail.com)

	% 1. An affiliation 2. A different affiliation {*}corresponding author(s):
	% Firstname Lastname (email@address)
	\begin{abstract}
		Singlet fission has emerged as one of the most exciting phenomena known to improve the efficiencies of different types of solar cells and has found uses in diverse optoelectronic applications. The range of available singlet fission molecules is, however, limited as to undergo singlet fission, molecules have to satisfy certain energy conditions. Recent advances in material search using inverse design has enabled the prediction of materials for a wide range of applications and has emerged as one of the most efficient methods in the discovery of suitable materials. It is particularly helpful in manipulating large datasets, uncovering hidden information from the molecular dataset and generating new structures. However, we seldom encounter large datasets in structure prediction problems in material science. In our work, we put forward inverse design of possible singlet fission molecules using a transfer learning based approach where we make use of a much larger ChEMBL dataset of structurally similar molecules to transfer the learned characteristics to the singlet fission dataset. 
	\end{abstract}
	
	\section*{Introduction}
	
	Singlet fission (SF) is a process by which a high-energy singlet exciton, resultant from the absorption of a photon, is converted into two triplet excitons, each carrying about half the energy. \cite{1, 2, 3, 4, 5} As the process is spin allowed, it can occur very rapidly (on a picosecond or femtosecond timescale) and out-compete radiative decay (that generally occurs on a nanosecond timescale) thereby producing two triplets with very high efficiency. \cite{1, 2} For a molecule to undergo SF, certain conditions need to be satisfied.  One of the main requirements is that the energy of the singlet state (E[S1]) should be more than double the energy of the triplet state (E[T1]), i.e., E[S1]/E[T1] $\geq$ 2. \cite{1,  2, 3} As an ideal case, E[S1]/E[T1] $\approx$ 2 for fast formation of the spin-coupled triplet state and too much exoergicity can result in unwanted heat generation. The other requirement for singlet fission is that the energy of the next higher triplet level should be greater than or equal to the energy of the two low-lying triplets, E[T2] $\geq$ 2$\times$E[T1]. \cite{1, 2, 3} SF offers the possibility of overcoming thermalisation losses in PVs, as every photon absorbed above the bandgap leads to the formation of two electron-hole pairs. Consequently, there is a growing scope of interest in this area as SF can overcome the theoretical Shockley–Queisser limit for the power conversion efficiency of a single junction solar cell, which is about 32\%.  \cite{6} However, the realization of SF-based solar cells is hindered by dearth of suitable materials. Due to the high computational cost of excited-state quantum mechanical calculations, predictive descriptors that are fast to evaluate must be found in order to explore the chemical space in search of new SF materials. \\ \\
	Previously, quantum mechanical methods and time dependent density functional theory (TD-DFT) have been used to the study of singlet fission process. Berkelbach et al. applied Redfield theory to investigate the charge transfer states in the SF dynamics of pentacene. \cite{4, 5, 7} Tamura et al. utilized the time-dependent Hartree quantum mechanical approach to study the SF mechanism in a pentacene derivative and in rubrene. \cite{8} Krishnapriya et.al. employed spin density distribution to encode SF in a series of pentacene dimers using phenyl-, thienyl- and selenyl- flanked diketopyrrolopyrrole (DPP) derivatives. \cite{9} Several research groups have also applied machine learning methods to study SF dynamics. Chen et.al. have reported that deep learning can be used for describing nonadiabatic excited state dynamics for SF. \cite{10} Schröder et.al. have used tree-tensor network state simulations to compute the real-time dynamics of exponentially large vibronic states of SF molecules. \cite{11} Although these interesting studies have explored the energetics and charge carrier dynamics of different existing SF molecules, these methods cannot predict new potential materials for SF. 
	Generally, for the development of new materials, the stepwise procedure of selection of materials, prediction of material properties, chemical synthesis, and experimental validation is usually repeated until satisfactory performance is achieved. As this experimental approach is tedious and expensive, progressively predictive techniques, for example, high-throughput computational screening (HTCS) have gained popularity in recent years. \cite{12, 13, 14, 15} In HTCS, extensive property prediction using DFT calculations \cite{16} or machine learning  \cite{17, 18} is carried out after screening out suitable materials from molecular libraries and open databases. This permits highly efficient categorization of potential candidates for subsequent experimental verification. This procedure has been increasingly used in materials science for various applications such as learning the chemistry of materials using only elemental composition \cite{35} , crystal structure prediction \cite{32} and target property prediction \cite{33, 34} . But for DFT calculations, the computational complexity increases with the number of atoms and achieving the results up to a certain accuracy can be computationally quite costly. Predicting material properties using machine learning requires the selection of suitable features and feature selection is intuitive as it cannot be reasoned out why some features work better than the others. Moreover, there is no assurance regarding whether the correct chemical space is being investigated. \\ \\
	To overcome the above-mentioned issues, inverse design of materials using deep learning has emerged as one of the most promising methods in recent years for predicting potential materials with specific target properties for an application.  \cite{26, 27, 28, 29, 30} Inverse design aims to design materials that are expected to meet the given target properties in a direct manner, whereas the conventional approach designs materials first and predicts their properties subsequently. The inverse design approach extracts the molecular design knowledge hidden in the molecular database and generates new molecules on the basis of its own knowledge, thereby allowing systematic materials exploration without the need for researcher experience or intuition. Deep learning has recently found significant use in the inverse design of molecules, especially in the area of drug discovery. Gomez-Bombarelli et al. \cite{19} have shown that deep generative models could be utilized for the inverse design of potential drug molecules by optimizing certain properties of importance. They used a Variational Auto-Encoder (VAE) \cite{20} , a robust generative model for this purpose, the architecture and specifics of which have been elaborated in later sections. Similarly, Popova et al. \cite{21} experimented with stack-augmented RNNs to design drug molecules, while using Reinforcement Learning approaches to tune properties such as solubility. \\ \\
	Our primary goal was to investigate the possibilities of inverse design of singlet fission molecules. The search for singlet fission molecules is hindered by the fact that there are too few molecules undergoing singlet fission. The dataset we used comprised of about 1000 screened singlet fission molecules from previous literature \cite{31, 1, 3, 5, 37, 38, 39, 40, 41}.\\ \\
	In our work, we propose an inverse design model for the generation of new potential singlet fission molecules. 
	Initially, we had applied a deep generative model to the 1000 molecule extract for the inverse design of singlet fission molecules. But, the number of molecules was not sufficient to successfully carry out inverse design using traditional deep learning techniques. This is because deep learning techniques try to represent the problem in the form of complex functions and therefore require large amounts of data to accurately describe the function. To solve this problem of limited data availability, we develop a new approach that makes use of a transfer learning approach. The transfer learning \cite{22} approach has primarily been applied to image classification and detection tasks in the past. Recently, this approach has started to gain recognition in molecular discovery field.  Segler et al. \cite{23}  used transfer-learning to
	first train the RNN on a whole dataset of molecules and later fine-tune the model towards the generation of drug molecules with physico-chemical properties of interest. Gupta et al. \cite{24} applied the transfer learning approach to grow drug molecules from fragments. In our approach, we initially train a deep network on a much larger dataset (ChEMBL - 100,000 molecules) \cite{25} to teach the model to encode and decode Simplified Molecular-Input Line-Entry System(SMILES) strings and understand the intricacies in their grammar. We then fine-tune this model on our singlet-fission dataset to transfer the learned characteristics to our task of importance. This fine-tuned model was then used to generate novel molecules that could possibly exhibit singlet fission property. The ChEMBL and singlet-fission datasets contained molecules that were structurally quite similar. This prompted us to choose the ChEMBL dataset for the pretraining task.

	\section*{Results}
	
	\subsection*{Computational Approach}
	
	\subsection*{Building the Generative Model}
	
\begin{figure}[H]
		\centering
		\includegraphics[width=0.8\textwidth]{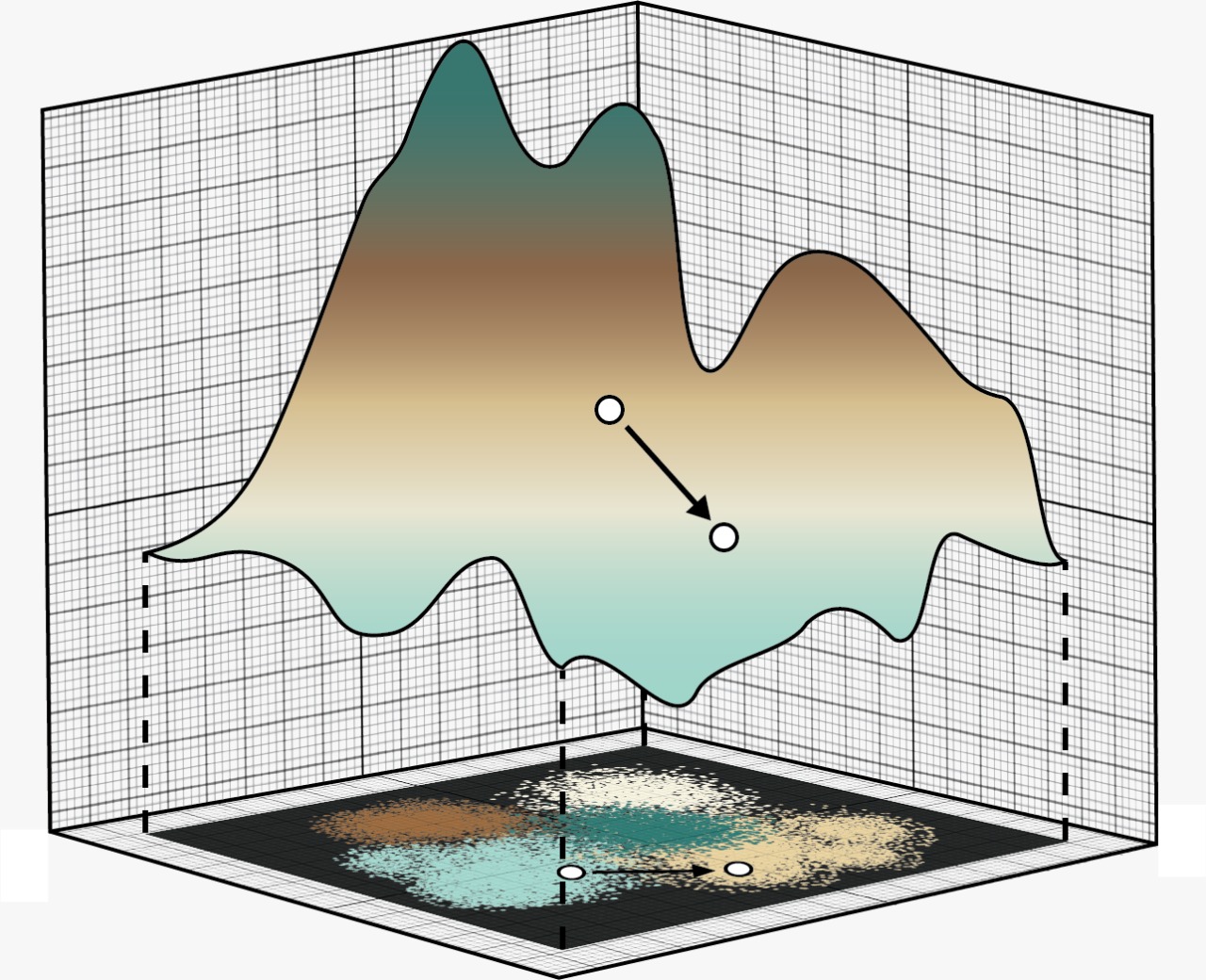}
		\caption{Latent space optimization. After the target property prediction of molecules based on their latent representation, we can optimize the values to find new latent representations expected to have better values of the desired properties.}
		\label{fig:latent}
	\end{figure}

	Our model primarily comprised of 3 networks : a) A Variational Autoencoder (VAE), b) A Multi Layer Perceptron (MLP) and c) A classifier.
	The VAE consists of an encoder that is responsible for encoding the input SMILES string into a 1-D vector, and a decoder that is responsible for decoding this latent vector back into a SMILES string. We call the 1-D encoded vector the latent representation of the molecule. A more detailed explanation of the autoencoder architecture is described in the Supplementary Information.
	For the generation of new structures, the chemical structures encoded in the continuous representation have to be connected with the target properties that we are seeking to optimize. We trained the VAE on a reconstruction task and the MLP on a regression task with the target property as E[S1] - 2 $\times$ E[T1], which we wish to optimize. 
	This results in the formation of a property-wise distribution of the latent space, which means that molecules with similar values of the target property will lie close to each other (measured by Euclidean distance) on the latent space. This then allows us to easily maneuver and explore the latent space, giving us flexibility while generating new molecular structures. A diagrammatic representation of the latent space is shown in Fig \ref{fig:latent}.
	
\begin{figure}[H]
		\centering
		\includegraphics[width=1.0\textwidth]{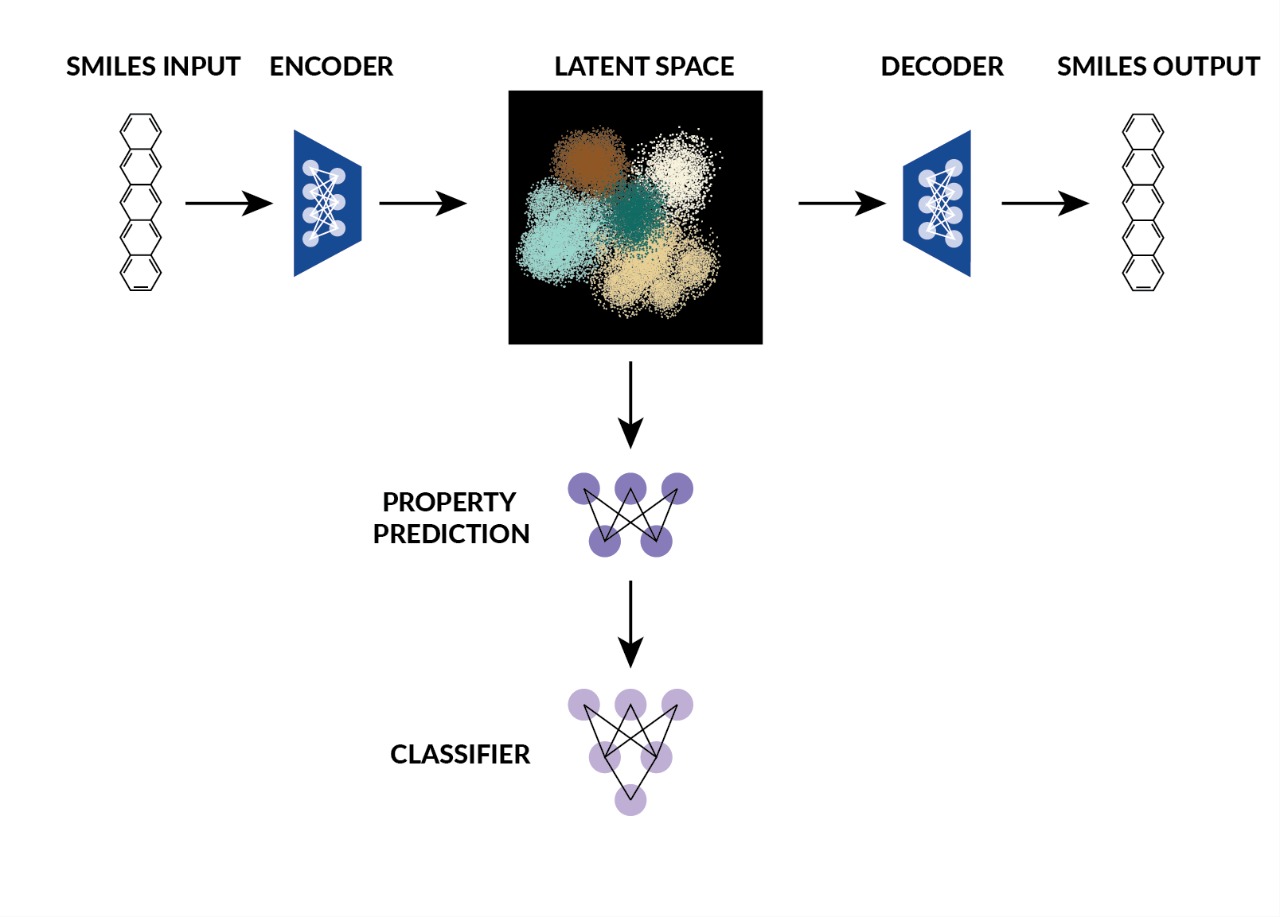}
		\caption{Flowchart for generating new structures using the inverse design model. The encoder converts the SMILES input into a 1-D vector which is decoded back to a SMILES string by the decoder. The formation of the property wise distribution of the latent space allows us to take the latent vector of an encoded molecule and move in a direction most likely to improve the target property, which in this case, is the first SF condition. For the second SF condition, a neural network is used as a binary classifier.}
		\label{fig:auto_flowchart}
	\end{figure}

	To generate promising new singlet fission molecules, the latent vector of an encoded molecule is taken and then we advance in the direction anticipated to improve the desired target property. 
	A schematic flowchart depicting the autoencoder along with a property prediction network is shown in Fig \ref{fig:auto_flowchart}.
	
	The resulting new candidate vectors can then be decoded into corresponding molecules. Then, for the second singlet fission condition E[T2] - 2$\times$E[T1] $>$ 0 to be satisfied, we train another neural network for classification of the generated structures.

	\subsection*{Applying this to SF dataset}
	
	Since the singlet fission dataset has only about 1000 molecules, accurate reconstruction of new molecules is not feasible using deep inverse design, since deep learning approaches often require large amounts of data to perform well. When we tried training our deep network on the singlet fission dataset, the model was not able to learn the intricate and complex SMILES grammar and hence, generated many invalid SMILES strings. This is a major hurdle in applying modern inverse design approaches to tasks that contain less amounts of data. As a result, we propose a new approach for the design of new structures using transfer learning which allowed our model to perform far better and generate a substantial number of valid SMILES strings.
	
	\subsection*{Transfer Learning Approach}
	
	For the transfer learning approach, we need a database which contains similar structures as there are in the singlet fission. We choose the ChEMBL database of bioactive molecules which are very similar in structure with the molecules in the singlet fission dataset after which we extract around 100,000 molecules for the model. Then, we train the VAE on the 100,000 molecule extract of CHEMBL database to teach the model to learn basic SMILES grammar.
	
\begin{figure}[H]
		\centering
		\begin{subfigure}{.9\textwidth}
			\centering
			\includegraphics[width=.9\textwidth]{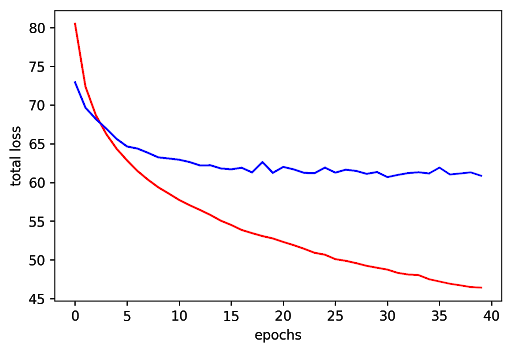}
			\caption{Plot of total training loss (reconstruction + regression) vs number of epochs}
			\label{fig:plot1}
		\end{subfigure}
		\begin{subfigure}{.9\textwidth}
			\centering
			\includegraphics[width=.9\textwidth]{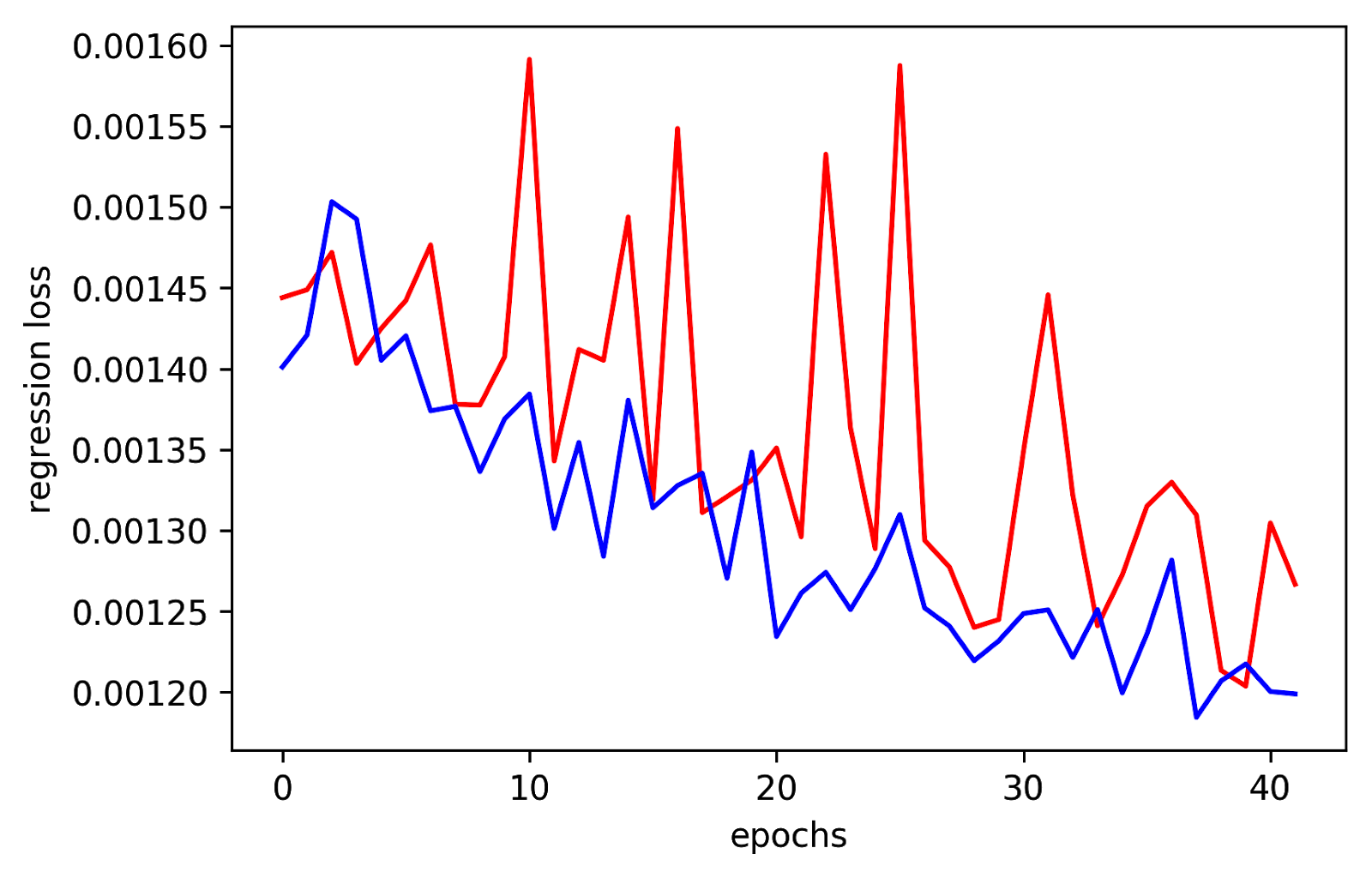}
			\caption{ Plot of regression loss vs number of epochs}
			\label{fig:plot2}
		\end{subfigure}
		\caption{ Training plots on the singlet fission dataset. Blue line indicates loss on the validation dataset and red line indicates loss on the training dataset}
		\label{fig:plots}
	\end{figure}

	We followed the following steps to transfer the learned characteristics from the ChEMBL dataset to the singlet fission dataset: 1) Pretrained the VAE on ChEMBL extract on only reconstruction of input SMILES strings, 2) Joint training of pretrained VAE and newly initialized MLP on Reconstruction + Regression (Property Prediction) tasks, 3) Froze VAE layers and only trained MLP on regression task (Model started overfitting on reconstruction task earlier than on the regression task during joint training).
	The loss curves obtained while training the model on the singlet fission dataset are shown in Fig \ref{fig:plots}.
	
	Our VAE contains Dense (Fully connected) layers not only as the final layer but also in some positions in the middle of the network. These dense layers are input size dependent and so we had to follow identical preprocessing methods in both the ChEMBL and Singlet Fission datasets so as to have identical input sizes in both cases. The presence of these size dependent layers was also the key reason that we chose to use the combined vocabulary of both datasets to define our fully connected layer sizes.
	
	\subsection*{Generating novel structures}
	
\begin{figure}[H]
		\centering
		\includegraphics[width=1.0\textwidth]{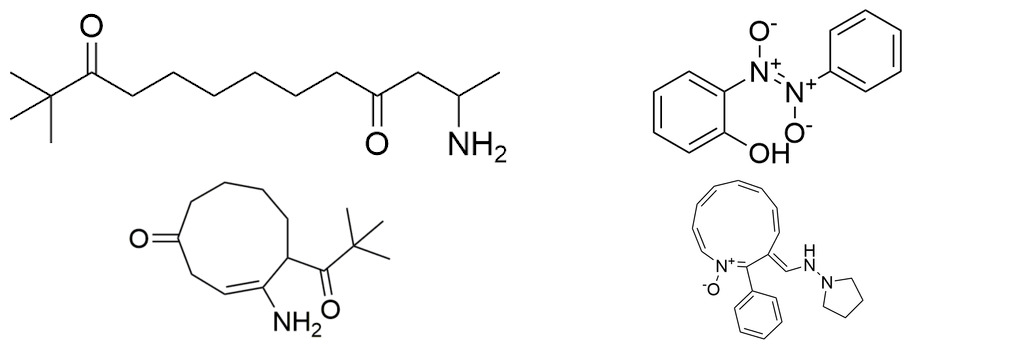}
		\caption{Generated molecular structures}
		\label{fig:mol_joint}
	\end{figure}

	To generate potential singlet fission molecules using our model, we decoded points in the latent space close (measured by Euclidean distance) to molecules already known to satisfy the Singlet Fission condition (E[S1] - 2 $\times$ E[T1] $>$ 0).
	The VAE being probabilistic in nature, generates many invalid SMILES strings. To account for this stochasticity, we performed 200 decoding attempts per molecule so as to produce valid strings. From these decoding attempts, we usually got a prominent molecule and many others appeared with lower frequencies.
	By optimizing the Singlet Fission property (E[S1] - 2 $\times$ E[T1]), we were able to generate multiple potential molecules satisfying the above criterion. After that, we input the generated molecules through a neural network classifier for satisfying the second singlet fission condition E[T2] $\geq$ 2$\times$E[T1]. We show four of the resultant structures in Fig \ref{fig:mol_joint}.
	
	We see that although all four structures satisfy the singlet fission criteria, the third and fourth structures seem unstable due to the presence of 9 and 10 membered rings respectively. We can therefore shortlist the first two structures for further experimental validation or DFT analysis.

	\section*{Discussion}
	
	From the above results, we see that transfer learning can be a powerful tool in applying deep generative neural networks for the inverse design of materials, especially on small sized datasets. This approach is generalizable to a large number of materials science problems owing to the fairly common issue of small sized datasets. To the best of our knowledge, this is the first time that a transfer learning based approach is being applied to the inverse design of materials while optimizing properties of interest. This method is easily extendable to many other materials science problems with small dataset constraints. We can screen out the final potential molecules for further study and analysis. Consequently, this formulation of a transfer learning based inverse design framework is expected to minimize the efforts in the computational screening of feasible molecules in a boundless search space.

	\section*{Methods}
	
	\subsection*{Data Collection and Preprocessing}
	
	Pretraining was done on a 100,000 molecule extract of the ChEMBL database. The learned features were then transferred to the singlet fission dataset \cite{31}.
	First, we take the molecules from the datasets and convert them into their SMILES representation by using RDKIT. Then, for feeding into the neural network, we generate one-hot encoded vector representations of the SMILES strings with a vocabulary (set of unique characters) of length 52. We chose this length by calculating the vocabulary of the combined singlet fission dataset + ChEMBL extract. For computational ease, we encoded strings up to a maximum length of 120 characters and padded shorter strings with spaces. We chose the above form of representation to allow for easy encoding of SMILES to vectors and decoding of vectors back to SMILES or in other words, a two-way mapping between the strings and their corresponding vectors.

	\subsection*{Model Architecture}
	
	We made many architectural design choices for our VAE based on work by Gomez-Bombarelli et al. The following was the construction of our VAE model: The encoder consisted of four 1-D convolutional layers of size 9, 9, 11 and 11 followed by two fully connected (Dense) layers of sizes 435 and 292 respectively. Our latent representation was therefore a 1-D vector of length 292. The decoder consisted of three Gated Recurrent Unit (GRU) layers of hidden size 501 followed by a fully connected layer of size 52 whose outputs were passed to a softmax activation function. Our decoder therefore produced a probability distribution over each character in our vocabulary for each of the 120 characters of our output string. Rectified Linear Unit activation function was applied after all hidden layers except for the pre-final layer of the encoder and the first layer of the decoder where Scaled Exponential Linear Unit (SELU) activation function was used. Adam was the optimizer used during the VAE training.
	Pretraining of the VAE on the reconstruction task on the ChEMBL extract required a total of 136 epochs to train and finetuning of VAE + MLP on joint reconstruction + regression (property prediction) tasks was carried out for 42 epochs after which our model started overfitting on the reconstruction task. This called for the third step of our training procedure which was to freeze all VAE layers and only train the MLP on the property prediction task for another 50 epochs to achieve the best possible results on both tasks.
	We used a learning rate of 1e-3 during pretraining of the VAE and a learning rate of 1e-4 during finetuning of the VAE + MLP model. All our experiments were carried out using the PyTorch \cite{36} framework.

	\section*{Data Availability}
	
	Data for training the model is available upon request.

	\section*{Code Availability}
	An implementation of the algorithm described in the paper and pretrained model weights are available at https://github.com/aksub99/FissionNet.

	\section*{Author Information}
	
	\subsection*{Affiliations}
	\textit{Department of Physics, Indian Institute of Technology Roorkee, Roorkee, Uttarakhand, 247667, India} \\
	Utkarsh Saha, Tejasvini Sharma, Naveen K. Tailor \& Soumitra Satapathi \\
	\\
	\textit{Department of Metallurgical and Materials Engineering, Indian Institute of Technology Roorkee, Roorkee, Uttarakhand, 247667, India} \\
	Akshay Subramanian

	\section*{Authors' contributions}
	
	S.S., A.S. and U.S. conceived the project, A.S. wrote the computer software and carried out simulations and experiments with contributions from U.S.; U.S., A.S., S.S., T.S. and N.K.T. wrote the manuscript. A.S. and U.S. contributed equally on this work.

	\section*{Competing interests}
	
	The authors declare no competing interests.
	
	\bibliographystyle{ScienceAdvances}
	\bibliography{publication}
\end{document}